\documentclass{svproc}
\usepackage[numbers,sectionbib]{natbib}
\usepackage{url}

\usepackage{amsmath}
\usepackage{algorithm}
\usepackage{algorithmicx}
\usepackage{amsmath,algorithm,tabularx}
\usepackage{algpseudocode}
\usepackage{amssymb}
\usepackage{multirow}
\usepackage{makecell}
\usepackage{caption}
\usepackage{array}
\usepackage[pdftex]{graphicx}

\newcolumntype{P}[1]{>{\centering\arraybackslash}p{#1}}
\newcolumntype{C}{>{\centering\arraybackslash} m{6cm} }
\makeatletter
\newcommand{\multiline}[1]{%
	\begin{tabularx}{\dimexpr\linewidth-\ALG@thistlm}[t]{@{}X@{}}
		#1
	\end{tabularx}
}
\makeatother

\begin{document}
\mainmatter              % start of a contribution
\title{Channel Type Recognition in Wireless Communications: A Deep Learning Approach}
\titlerunning{Channel Type Recognition}  % abbreviated title (for running head)

\author{Shu Sun, Xiaofeng Li, Sungho Moon}
\authorrunning{Channel Type Recognition} % abbreviated author list (for running head)
\institute{}

\maketitle              % typeset the title of the contribution

\begin{abstract}
In this paper, we propose two novel and practical deep-learning-based algorithms to solve the wireless channel type (WCT) recognition problem. Specifically, the WCT recognition problem is recast as a classification problem in deep learning due to their similarities, where a deep neural network (DNN) is trained offline with a diversity of typical WCTs for fifth-generation (5G) and beyond-5G wireless communications, which is then utilized to perform online WCT determination. In the first algorithm, one WCT is regarded as a single task. While in the second scheme, one WCT is jointly characterized by several independent features, each of which is treated as a task and is classified respectively by training a DNN in a multi-task-learning manner, and the final WCT is identified by the combination of those channel features. Simulation results show that the proposed algorithms can classify various WCTs instantaneously with high accuracy, result in satisfactory block error rate and throughput, and outperform a representative baseline WCT determination scheme. 
\keywords{classification, deep learning, deep neural network (DNN), wireless channel type}
\end{abstract}
\section{Introduction}
In wireless communications, data should be transmitted using a proper modulation and coding scheme (MCS) depending on the channel conditions in order for the receiver (Rx) to be able to decode the data with a target bit error rate (BER) or block error rate (BLER). Which MCS to use is usually determined based upon estimated channel state information (CSI) feedback obtained using CSI reference signals (CSI-RSs) for the downlink, sounding reference signals (SRSs) for the uplink, or other types of reference signals as per the 3rd Generation Partnership Project (3GPP) specifications for the fifth-generation (5G) and beyond-5G (B5G) wireless systems~\cite{38211,38213,38214,38331}. The CSI estimation typically includes channel matrix estimation, signal-to-noise ratio (SNR) estimation and/or mutual information (MI) estimation, as well as MCS estimation. The mapping to MCS from prior estimated parameters, such as SNR and/or MI, often relies on the wireless channel type (WCT) to ensure that a target BER or BLER can be fulfilled. For instance, if the MCS is mapped from the MI, then multiple MI-to-MCS mapping tables~\cite{Sun20} are indispensable for various WCTs, such as additive white Gaussian noise (AWGN) channel, EPA5 (Extended Pedestrian A model with 5 Hz Doppler frequency) with low spatial correlation between Rx antennas, EVA5 (Extended Vehicular A model with 5 Hz Doppler frequency) with high spatial correlation between Rx antennas, among others. More importantly, the mapping table can vary significantly for different WCTs due to their distinct channel properties. Therefore, it is critical to identify the WCT in order to select the correct mapping table. 

WCT recognition is challenging since a wireless channel involves multiple entangled characteristics including Doppler spread, spatial correlation, frequency selectivity or equivalently delay spread, among others, which implies that WCT recognition necessitates the identification of all those channel properties. Moreover, many of the channel properties may take continuous values in reality, while quantization or several thresholds are needed in practical hardware and/or firmware implementations to facilitate signal processing, where the quantization precision or the threshold values themselves need further and judicious determination. Consequently, it is prohibitively complicated to determine the WCT based upon conventional human effort. On the other hand, deep learning, an important branch of artificial intelligence (AI), has found numerous applications in wireless communications including the physical layer~\cite{Li17WC,O'Shea17TCCN,Raj18CL,Nawaz19Access,Huang20WC,Sun20Fdra}. Specifically, deep learning has been employed to perform automatic modulation classification~\cite{West17DySPAN,Rajendran18TCCN,Soltani19MILCOM}, signal detection~\cite{Ye18WCL}, beamspace channel estimation~\cite{He18WCL}, beam selection~\cite{Li19Asilomar,Sim20Access}, multiple-input-multiple-output (MIMO) detection~\cite{Samuel19TSP}, power allocation~\cite{Yan20Access,Qian20Access,Yan20IoT,Yan20Globecom}, and resource allocation~\cite{Sun21DrlFdra}, to name a few. 
To the authors' best knowledge, however, there has been no well-recognized publicly available solution to WCT determination (either with or without AI techniques). Deep learning based end-to-end communication systems have been demonstrated in~\cite{Qin19WC}. Note that, however, in practical wireless deployment in the industry, it is preferred to carry out block-by-block (or module-by-module) wireless system processing, instead of in an end-to-end manner, especially when utilizing AI techniques, in order to enhance the tractability and to facilitate the debugging and maintenance of the system. Therefore, it is desirable and necessary to design an intelligent WCT recognition functionality block to pave the way for subsequent Rx procedures in a wireless communication system. 

In this paper, we propose two novel approaches to WCT recognition leveraging deep learning techniques. Specifically, we recast the WCT determination problem as a classification problem in deep learning, where a deep neural network (DNN) is first trained offline with a variety of WCTs that are representative in 5G and B5G wireless communications, and is then employed to conduct online WCT recognition. In particular, in the first approach, one WCT is regarded as one task, which is classified using a DNN trained offline in a single-task learning manner. While in the second algorithm, a WCT is jointly characterized by delay spread/frequency selectivity, spatial correlation between Rx antennas, Doppler spread, and so on, each of which is treated as a task and is classified separately by training a DNN via multi-task learning~\cite{Zhang17NSR}, and the final WCT is determined by the combination of those channel features, where the learned features can be employed for other purposes. 

\section{System Model}
We consider a wireless communication link with one next-generation nodeB (gNB) and one user equipment (UE), where the gNB and UE are equipped with $N_\text{G}$ and $N_\text{U}$ antennas, respectively, where $N_\text{G},N_\text{U}\geq1$. The gNB and UE communicates with each other through a wireless channel, and the transmitted signal is contained in $B$ resource blocks (RBs) and $M$ symbols, where an RB consists of 12 consecutive sub-carriers~\cite{38211}. At the Rx side (UE for downlink and gNB for uplink), the received signal is first transformed into the frequency domain via FFT, followed by descrambling, which inherently carries channel features and are free from channel estimation errors since they are obtained prior to channel estimation. Therefore, the descrambled signal at the Rx can be employed to assess the WCT. 

\section{Proposed WCT-Recognition Algorithms Based on Deep Learning}
In this section, we elaborate on the proposed two deep-learning-enabled WCT recognition algorithms. In the first algorithm, the WCT recognition problem is formulated as a single-task classification problem, where the task is to directly categorize the unknown WCT into one of the types used for training the DNN. In the second algorithm, the original problem is divided into several tasks, each of which represents a feature of the wireless channel to be determined and is classified respectively by training a DNN via multi-task learning, and the final WCT is determined by the combination of those channel characteristics. 

\subsection{Single-Task Classification}
Without loss of generality, for a given WCT, transmission slot, and SNR combination, the total length of the descrambled signal sequence over all RBs, symbols, and Rx antennas is denoted as $N_\text{des}$. The steps for using single-task classification to determine the WCT are detailed in Algorithm 1. Specifically, the descrambled signals in all the RBs, symbols, and Rx antennas used are rearranged into a vector by separating the real and imaginary parts of each descrambled complex signal, putting the real part of all descrambled signals into a vector, putting the imaginary part of all descrambled signals into a vector, and concatenating the two vectors into one. The procedure above is repeated for a wide range of SNR values to eliminate SNR dependency during WCT detection, for a large number of slots to obtain statistically sufficient samples, and for all potential WCTs that are of interest. Afterwards, the training and inference data sets are labeled where one unique label is created for each of the WCTs considered, as demonstrated by Step 12 in Algorithm 1. For example, assuming the WCTs considered are AWGN, EPA5 with low correlation, EPA5 with high correlation, EVA700 with low correlation, and EVA700 with high correlation, then the labels for the five WCTs can be set to the one-hot vectors [1 0 0 0 0], [0 1 0 0 0], [0 0 1 0 0], [0 0 0 1 0], and [0 0 0 0 1], respectively. Ultimately, the aforementioned training and inference data sets are employed to train a DNN offline to obtain a trained DNN that can classify an instantaneous WCT online. 

\begin{algorithm}
	\caption{Single-Task Classification}
	\begin{algorithmic}[1]% The number tells where the line numbering should start
		\For{$n_\text{WCT}=0:N_\text{WCT}-1$}
		\For{$n_\text{slot}=0:N_\text{slot}-1$}
		\For{$n_\text{SNR}=0:N_\text{SNR}-1$}
		\State \multiline{%
			$\textbf{s}_{n_\text{WCT},n_\text{slot},n_\text{SNR}}=\textbf{a}_{n_\text{WCT},n_\text{slot},n_\text{SNR}}+j\textbf{b}_{n_\text{WCT},n_\text{slot},n_\text{SNR}}$, where $\textbf{s}_{n_\text{WCT},n_\text{slot},n_\text{SNR}}\in\mathbb{C}^{N_\text{des}\times1}$ denotes the descrambled signal sequence and $\textbf{a}_{n_\text{WCT},n_\text{slot},n_\text{SNR}},\textbf{b}_{n_\text{WCT},n_\text{slot},n_\text{SNR}}\in\mathbb{R}^{N_\text{des}\times1}$}
		\State \multiline{%
			$\tilde{\textbf{a}}_{n_\text{WCT},n_\text{slot},n_\text{SNR}}=\left[\textbf{a}_{n_\text{WCT},n_\text{slot},n_\text{SNR}};\textbf{b}_{n_\text{WCT},n_\text{slot},n_\text{SNR}}\right]\in\mathbb{R}^{2N_\text{des}\times1}$}  
		\EndFor
		\EndFor
		\EndFor
		\State \multiline{%
			$\textbf{S}=\left[\tilde{\textbf{a}}_{0,0,0},...,\tilde{\textbf{a}}_{N_\text{WCT}-1,N_\text{slot}-1,N_\text{SNR}-1}\right]\in\mathbb{R}^{2N_\text{des}\times N_\text{WCT}N_\text{slot}N_\text{SNR}}$}  
		\State \multiline{%
			Obtain training data set: $\textbf{S}_\text{training}=\left[\textbf{S}\right]_{[:,0:\alpha N_\text{WCT}N_\text{slot}N_\text{SNR}-1]}\in\mathbb{R}^{2N_\text{des}\times\alpha N_\text{WCT}N_\text{slot}N_\text{SNR}}$, where $0<\alpha<1$}  
		\State \multiline{%
			Obtain inference data set: $\textbf{S}_\text{inference}=\left[\textbf{S}\right]_{[:,\alpha N_\text{WCT}N_\text{slot}N_\text{SNR}:N_\text{WCT}N_\text{slot}N_\text{SNR}-1]}\in\mathbb{R}^{2N_\text{des}\times(1-\alpha)N_\text{WCT}N_\text{slot}N_\text{SNR}}$}  
		\State \multiline{%
			Label the training and inference data: create an $N_\text{WCT}\times N_\text{WCT}$ identity matrix $\textbf{I}_{N_\text{WCT}\times N_\text{WCT}}$, then the label for the $n_\text{WCT}$-th WCT is the one-hot vector $\textbf{e}=\left[\textbf{I}_{N_\text{WCT}\times N_\text{WCT}}\right]_{\left[:,n_\text{WCT}\right]}\in\mathbb{R}^{N_\text{WCT}\times1}$}  
		\Comment{The dimensions of the labels for training and inference data sets are $N_\text{WCT}\times\alpha N_\text{WCT}N_\text{slot}N_\text{SNR}$ and $N_\text{WCT}\times(1-\alpha)N_\text{WCT}N_\text{slot}N_\text{SNR}$, respectively}
		\State \multiline{%
			Use the aforementioned training data set and inference data set to train a DNN offline to obtain a trained DNN that can determine an instantaneous WCT online}
	\end{algorithmic}
\end{algorithm}

\subsection{Multi-Task Classification}
The detailed steps for multi-task classification are provided in Algorithm 2, where the method for generating the training and inference data sets is the same as Steps 1-11 in Algorithm 1. The key discrepancy between single-task and multi-task classifications lies in how the training and inference data sets are labeled. As pointed out previously, a WCT can be jointly characterized by a few features including delay spread/frequency selectivity, channel correlation type, and Doppler spread, among others. Let's take delay spread, channel correlation type, and Doppler spread these three features as an example. Assuming the WCT considered are AWGN with low correlation (and 0 Hz Doppler Spread), EPA5 with low correlation, EPA5 with high correlation, EVA700 with low correlation, and EVA700 with high correlation, then there are three frequency selectivity categories (AWGN, EPA, and EVA), two spatial correlation types (low and high), and three Doppler spread values (0, 5, and 700 Hz). The labels for the three frequency selectivity categories are [1 0 0], [0 1 0], and [0 0 1], the labels for the two spatial correlation types are [1 0] and [0 1], and the labels for the three Doppler spread values are [1 0 0], [0 1 0], and [0 0 1]. Finally, the labels for each characteristic are concatenated into an $8\times1$ vector to form the overall label for multi-task (three-task in this case) learning using the DNN, e.g., EPA5 with high correlation corresponds to the label [0 1 0 0 1 0 1 0], where the first three digits [0 1 0] indicates EPA, the following two digits [0 1] represents high correlation, and the last three digits [0 1 0] denotes 5 Hz Doppler spread. The labels for the other WCTs can be derived similarly. Then the training and inference data sets generated above are employed to train a DNN offline to obtain a trained DNN that can determine an instantaneous WCT online, where the WCT is determined via the combination of all the channel features assorted by the DNN. 

It is noteworthy that in addition to being utilized for WCT recognition, each of the wireless channel features acquired from the multi-task learning can also be adopted individually in subsequent Rx processing. 
\begin{algorithm}
	\caption{Multi-Task Classification}
	\begin{algorithmic}[1]% The number tells where the line numbering should start
		\State \multiline{%
			Generate training and inference data sets: identical to Steps 1-11 in Algorithm 1}  
		\Comment{Labeling of training and inference data begins}
		\State \multiline{%
			Among all the WCTs to be classified, find the numbers of distinct delay spread, channel correlation, and Doppler spread types, denoted as $N_\text{DS}$, $N_\text{corr}$, and $N_\text{Dopp}$, respectively.}  
		\State \multiline{%
			Create three identity matrices $\textbf{I}_{N_\text{DS}\times N_\text{DS}}$, $\textbf{I}_{N_\text{corr}\times N_\text{corr}}$, and $\textbf{I}_{N_\text{Dopp}\times N_\text{Dopp}}$, then the labels for the $n_\text{DS}$-th delay spread, $n_\text{corr}$-th correlation type, and $n_\text{Dopp}$-th Doppler spread are the one-hot vectors $\textbf{e}_\text{DS}=\left[\textbf{I}_{N_\text{DS}\times N_\text{DS}}\right]_{\left[:,n_\text{DS}\right]}\in\mathbb{R}^{N_\text{DS}\times1}$, $\textbf{e}_\text{corr}=\left[\textbf{I}_{N_\text{corr}\times N_\text{corr}}\right]_{\left[:,n_\text{corr}\right]}\in\mathbb{R}^{N_\text{corr}\times1}$, and $\textbf{e}_\text{Dopp}=\left[\textbf{I}_{N_\text{Dopp}\times N_\text{Dopp}}\right]_{\left[:,n_\text{Dopp}\right]}\in\mathbb{R}^{N_\text{Dopp}\times1}$, respectively}  
		\State \multiline{%
			The label for the WCT comprising the $n_\text{DS}$-th delay spread, $n_\text{corr}$-th correlation type, and $n_\text{Dopp}$-th Doppler spread is $\textbf{e}=\left[\textbf{e}_\text{DS};\textbf{e}_\text{corr};\textbf{e}_\text{Dopp}\right]\in\mathbb{R}^{\left(N_\text{DS}+N_\text{corr}+N_\text{Dopp}\right)\times1}$}
		\Comment \multiline{%
			The dimensions of the labels for training and inference data sets are $\left(N_\text{DS}+N_\text{corr}+N_\text{Dopp}\right)\times\alpha N_\text{WCT}N_\text{slot}N_\text{SNR}$ and $\left(N_\text{DS}+N_\text{corr}+N_\text{Dopp}\right)\times(1-\alpha)N_\text{WCT}N_\text{slot}N_\text{SNR}$, respectively \\ Labeling of training and inference data ends}
		%\Comment{Labeling of training and inference data ends}
		\State \multiline{%
			Use the aforementioned training data set and inference data set to train a DNN offline to obtain a trained DNN that can determine an instantaneous WCT online, where the WCT is determined via the combination of all the channel features classified by the DNN}
	\end{algorithmic}
\end{algorithm}

\subsection{DNN Architectures}
\begin{figure}
	\centering
	\includegraphics[width=0.8\columnwidth]{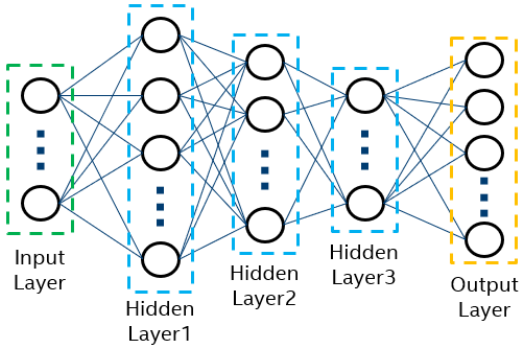}
	\caption{An example of single-task DNN architecture used in this work.}
	\label{fig:DNN1}	
\end{figure}

\begin{figure}
	\centering
	\includegraphics[width=0.8\columnwidth]{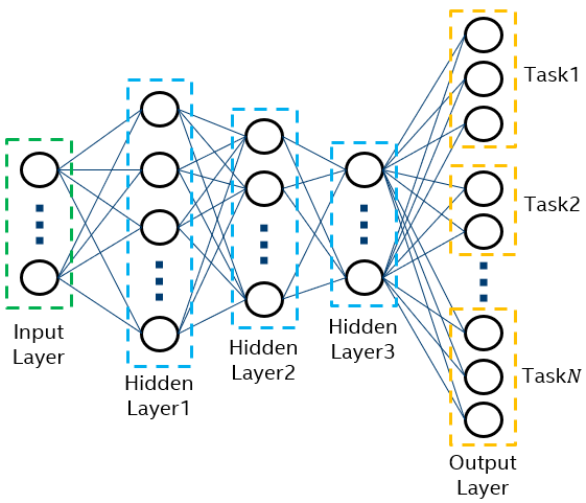}
	\caption{An example of multi-task DNN architecture used in this work.}
	\label{fig:DNN2}	
\end{figure}

A paradigm of the DNN employed in this work is composed of five fully-connected layers: one input layer, three hidden layers, and one output layer. The dimension of the input layer is $2N_\text{des}\times1$. The numbers of neurons in the three hidden layers are 3500, 2000, and 500, respectively, which are obtained based on joint optimization with a series of other parameters such as the number of input samples which is in turn determined by the amounts of SNRs, slots, WCTs, as well as training and inference data, which will be detailed in the next section. Fig.~\ref{fig:DNN1} illustrates an example of the single-task DNN used in our work, in which the dimension of the output layer is $N_\text{WCT}\times1$. An example of the multi-task DNN is depicted in Fig.~\ref{fig:DNN2}, where the output layer contains the labels for multiple tasks and has a dimension of $\sum_{n_\text{fea}=1}^{N_\text{fea}}K_{n_\text{fea}}\times1$, where $N_\text{fea}$ is the total number of wireless channel features considered, and $K_{n_\text{fea}}$ indicates the number of unique values for the $n_\text{fea}$-th wireless channel feature.
\begin{table}[!t]
	\renewcommand{\arraystretch}{1.2}
	\caption{Simulation settings for WCT recognition using DNN.}
	\label{tbl:simSet}
	\centering
	\begin{tabular}{|c||c|}
		\hline
		\textbf{Parameter} & \textbf{Value} \\
		\hline \hline
		\makecell{Number of SRS symbols}& \makecell{2} \\
		\hline
		\makecell{Number of RBs}& \makecell{16} \\
		\hline
		\makecell{SRS comb~\cite{38211,38214}}& \makecell{2} \\
		\hline
		\makecell{Number of Rx antennas}& \makecell{2} \\
		\hline
		\makecell{Dimension of each input sample}& \makecell{$768\times1$} \\
		\hline
		\makecell{Number of SNRs}& \makecell{31 (0 dB to 30 dB \\in increments of 1 dB)} \\
		\hline
		\makecell{Number of slots per SNR}& \makecell{500} \\
		\hline
		\makecell{Number of WCTs}& \makecell{5} \\
		\hline
		\makecell{WCTs}& \makecell{AWGN, \\ EPA5 low correlation, \\ EPA5 high correlation, \\EVA5 low correlation, \\EVA5 high correlation} \\
		\hline
		\makecell{Acceptable BLER}& \makecell{Within 10\%} \\
		\hline
	\end{tabular}
\end{table}
\begin{table}[!t]
	\renewcommand{\arraystretch}{1.5}
	\caption{Simulation parameters of the DNN}
	\label{tbl:drlSimSet}
	\centering
	\begin{tabular}{|c||c|}
		\hline
		Parameter & Value \\
		\hline \hline
		Number of hidden layers & 3 \\
		\hline
		Number of neurons per hidden layer & 3500, 2000, 500 \\
		\hline
		Initial weight value & Normal initialization \\
		\hline
		Optimization algorithm & Adam \\
		\hline
		Activation function & \makecell{First two layers: SELU\\ Last layer: Sigmoid}\\
		\hline
		Learning rate & 1e-3 \\
		\hline
		Batch size & 2048 \\
		\hline
	\end{tabular}
\end{table}

\section{Simulation Results}
To demonstrate the viability and effectiveness of the proposed DNN-based WCT recognition algorithms, we have performed simulations using the uplink SRS~\cite{38211,38214}. The simulation settings are given in Table~\ref{tbl:simSet}, where five representative WCTs are considered: AWGN, EPA5 with low correlation, EPA5 with high correlation, EVA5 with low correlation, and EVA5 with high correlation. Table~\ref{tbl:drlSimSet} lists the key DNN parameter values. The corresponding simulation results for a single-task DNN and a multi-task DNN are shown in Table~\ref{tbl:simRes1} and Table~\ref{tbl:simRes2}, respectively. It is evident from Table~\ref{tbl:simRes1} that the proposed single-task DNN is able to distinguish the given five WCTs with high accuracy (around 90\%). Furthermore, Table~\ref{tbl:simRes2} demonstrates that the proposed multi-task DNN can successfully categorize the channel features with high accuracy for all the tasks given, with an ultimate WCT classification accuracy of around 84\% taking into account all the channel features identified. In addition to the performance, it is noteworthy that the offline training time is relatively short based on our observations. For instance, with five WCTs and the data dimensions listed in Tables~\ref{tbl:simRes1} and~\ref{tbl:simRes2}, the training time is only about 15 minutes for both single-task and multi-task DNNs to achieve satisfactory classification accuracy as shown in those tables, and the online inference time is negligible as compared with the slot duration. 
\begin{table}[!t]
	\renewcommand{\arraystretch}{1.2}
	\caption{Simulation results for single-task DNN corresponding to the simulation settings in Table~\ref{tbl:simSet}}
	\label{tbl:simRes1}
	\centering
	\begin{tabular}{|c||c|}
		\hline
		\textbf{Task} & WCT \\
		\hline \hline 
		\makecell{\textbf{Input dimension of training data set}}& \makecell{$768\times69750$} \\
		\hline
		\makecell{\textbf{Output dimension of training data set}}& \makecell{$5\times69750$} \\
		\hline
		\makecell{\textbf{Input dimension of inference data set}}& \makecell{$768\times7750$} \\
		\hline
		\makecell{\textbf{Output dimension of inference data set}}& \makecell{$5\times7750$} \\
		\hline \hline 
		\textbf{Classification Accuracy} & 90.0\% \\
		\hline
	\end{tabular}
\end{table}
\begin{table*}
	\renewcommand{\arraystretch}{1.3}
	\caption{Simulation results for multi-task DNN corresponding to the simulation settings in Table~\ref{tbl:simSet}}
	\label{tbl:simRes2}
	\centering
	%\begin{tabular}{|c||c|c|c|}
	\begin{tabular}{|m{3.7cm}||m{2.5cm}|m{2.5cm}|m{2.5cm}|}
		\hline
		\makecell{\textbf{Task}} & \makecell{Delay spread\\/Frequency \\selectivity} & \makecell{Channel \\correlation} & \makecell{Doppler spread} \\
		\hline \hline 
		\textbf{\makecell{Input dimension \\of training data set}}& \multicolumn{3}{c|}{\makecell{$768\times69750$}} \\
		\hline
		\textbf{\makecell{Output dimension \\of training data set}}& \makecell{$3\times69750$} & \makecell{$3\times69750$} & \makecell{$2\times69750$}  \\
		\hline
		\textbf{\makecell{Input dimension \\of inference data set}}& \multicolumn{3}{c|}{\makecell{$768\times7750$}} \\
		\hline
		\textbf{\makecell{Output dimension \\of inference data set}}& \makecell{$3\times7750$} & \makecell{$3\times7750$} & \makecell{$2\times7750$} \\
		\hline \hline 
		\textbf{\makecell{Per-task \\classification accuracy}} & \makecell{95.0\%} & \makecell{88.6\%} & \makecell{99.9\%} \\
		\hline
		\textbf{\makecell{Overall \\classification accuracy}}& \multicolumn{3}{c|}{\makecell{84.1\%}} \\
		\hline
	\end{tabular}
\end{table*}

As mentioned previously, each WCT corresponds to a unique calibrated MI-to-MCS mapping table which reflects the maximum MCS level the channel can support given the SNR/MI and the BLER requirement (no larger than 10\% in this work). In other words, the WCT determines which MI-to-MCS table and hence which MCS to use for transmission, which in turn leads to an associated BLER and throughput. Therefore, to further demonstrate the effectiveness of the proposed WCT recognition method and to evaluate its impact on BLER and throughput, we have also performed simulations where the offline-trained multi-task DNN is employed for online WCT recognition, followed by the MI-to-MCS table selection based on the identified WCT and regular subsequent Rx processing to obtain BLER and throughput. Two baseline schemes are adopted for comparison: (1) perfect WCT knowledge, i.e., the WCT is assumed to be known a priory; (2) random WCT selection, where the WCT in each slot is randomly selected from the five candidate WCTs in Table~\ref{tbl:simSet}. 

\begin{figure}
	\centering
	\includegraphics[width=\columnwidth]{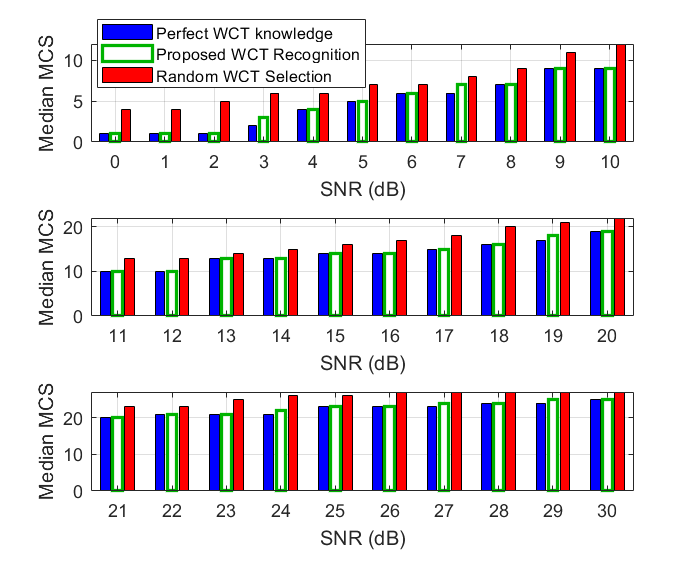}
	\caption{Median MCS with three different approaches: perfect WCT knowledge, proposed WCT recognition via multi-task learning, and random WCT selection.}
	\label{fig:mcs}	
\end{figure}
\begin{figure}
	\centering
	\includegraphics[width=0.9\columnwidth]{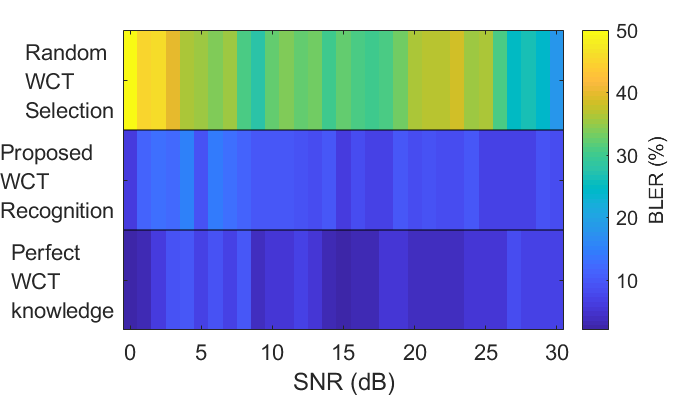}
	\caption{Block error rate (BLER) with three different approaches: perfect WCT knowledge, proposed WCT recognition via multi-task learning, and random WCT selection.}
	\label{fig:bler}	
\end{figure}
\begin{figure}
	\centering
	\includegraphics[width=0.9\columnwidth]{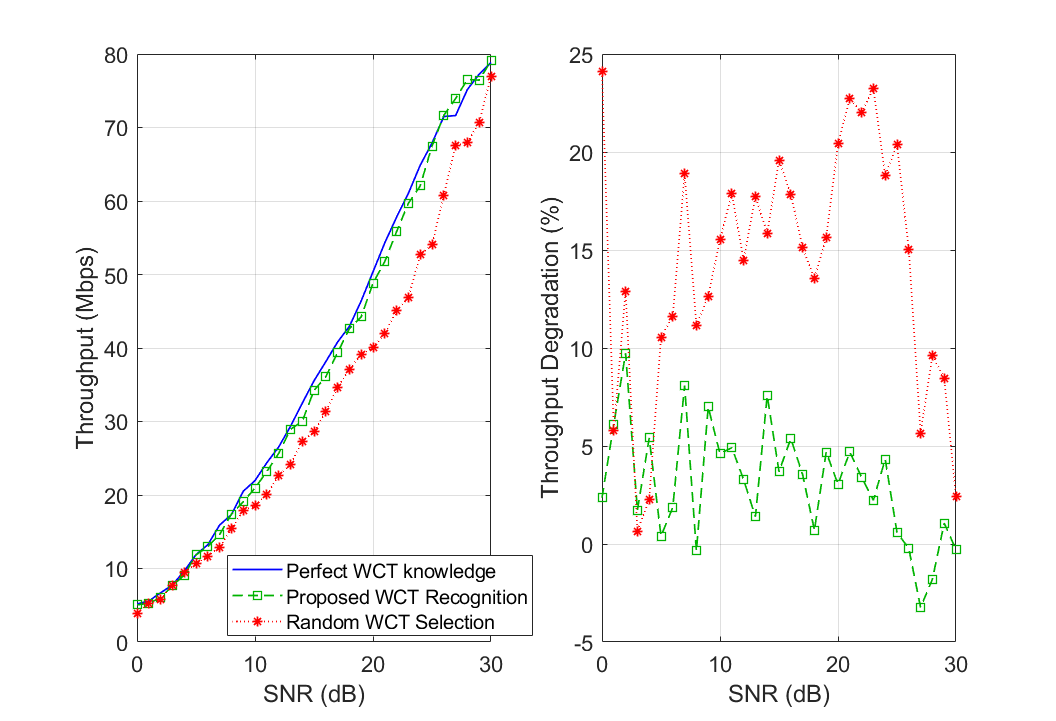}
	\caption{Throughput with three different approaches: perfect WCT knowledge, proposed WCT recognition via multi-task learning, and random WCT selection. The throughput degradation is with respect to the perfect WCT knowledge case.}
	\label{fig:thp}	
\end{figure}
 
 Fig.~\ref{fig:mcs} illustrates the median MCS performance versus the SNR, which shows that the proposed algorithm yields the same MCS as with perfect WCT knowledge in most cases, and the MCS difference is at most 1 between these two schemes. On the contrary, the random WCT selection method gives rise to higher MCS at all SNRs, where the largest MCS gap reaches 4 as compared with the former two approaches. The BLER performance is investigated in Fig.~\ref{fig:bler}, from which it can be observed that the BLER values corresponding to the perfect WCT case are between 1\% and 10\% for all SNRs as expected, thanks to the proper MCS acquired from the correct MI-to-MCS mapping table. Using the proposed algorithm, the majority of the BLER values are still maintained within 10\% with a maximum value of about 15\%. In contrast, the BLER values resulting from the random WCT selection scheme significantly exceed 10\% for all SNRs, and the highest even reaches 50\%, which is attributed to the excessively high MCS levels stemming from incorrect MI-to-MCS mapping tables as indicated by Fig~\ref{fig:mcs}. In Fig.~\ref{fig:thp}, we examine the throughput and throughput degradation with respect to the perfect WCT knowledge case. As evident from Fig.~\ref{fig:thp}, the proposed algorithm produces comparable throughput as compared with the perfect WCT knowledge case, where the throughput degradation is within 5\% for most SNRs. The random WCT selection scheme, however, generates substantially lower throughput in most cases with a maximum throughput degradation of approximately 25\%, owing to the high BLER values shown in Fig.~\ref{fig:bler}. 

\section{Conclusion}
In this paper, we have put forth two new and practical WCT recognition algorithms leveraging deep learning techniques, where the original WCT determination problem is formulated as a classification problem with single or multiple tasks. Simulation results show that the proposed algorithms can determine the WCT instantaneously with high accuracy, and yield satisfactory BLER values and throughput, while outperforming the random WCT selection baseline scheme. The proposed WCT recognition methods can save a large amount of time and energy spent on trial and error to characterize the WCT in field tests and deployment for next-generation wireless systems.

% ---- Bibliography ----
%
\bibliographystyle{my_spmpsci}  
\bibliography{DRL_Type-1_FDRA}

\end{document}